\documentstyle[psfig]{mn}
\newcommand{\kms}{\, {\rm km\, s}^{-1}}
\begin{document}
\def\lsim{\mathrel{\hbox{\rlap{\hbox{\lower4pt\hbox{$\sim$}}}\hbox{$<$}}}}
\def\gsim{\mathrel{\hbox{\rlap{\hbox{\lower4pt\hbox{$\sim$}}}\hbox{$>$}}}}
\def\simlt{\mathrel{\rlap{\lower 3pt\hbox{$\sim$}}
        \raise 2.0pt\hbox{$<$}}}
\def\simgt{\mathrel{\rlap{\lower 3pt\hbox{$\sim$}}
        \raise 2.0pt\hbox{$>$}}}

\title[Quasars and galaxy formation]
{Quasars and galaxy formation}
\author[Andrea Cattaneo]
{Andrea~Cattaneo $^1$\\
$^1$Institute of Astronomy, Madingley Road, Cambridge CB3 0HA}

\maketitle
\begin{abstract}

Quasars are widely believed to be powered by accretion onto supermassive
black holes and there is now considerable evidence for a link between 
mergers, the activity of quasars and the formation of spheroids.
Cattaneo, Haehnelt \& Rees (1999) have demonstrated that a very simple model 
in which supermassive black holes form and accrete most of their mass 
in mergers of galaxies of comparable masses can reproduce the observed
relation of black hole mass to bulge luminosity. Here we show that this 
simple model can account for the luminosity function of quasars and for the
redshift evolution of the quasar population provided a few additional 
assumptions are made. We use the extended Press-Schechter formalism to
simulate the formation of galaxies in hierarchical models of the formation
of structures and we assume that, when two galaxies of comparable masses merge,
their central black holes coalesce and a fraction of the gas in the
merger remnant is accreted by the supermassive black hole over a time-scale
of $\sim 10^7\,$yr. We find that the decrease in the merging rate with cosmic
time and the depletion in the amount of cold gas available due to the
formation of stars are not sufficient to explain the strong decline in the
space density of bright quasars between $z=2$ and $z=0$, since larger and
larger structures form, which can potentially host brighter and brighter
quasars. To explain the redshift evolution of the space density
of bright quasars
in the interval $0<z<2$ we need to assume that there is a dependence on
redshift either in the fraction of available gas accreted or 
in the time-scale for accretion. 
\end{abstract}

\section{Introduction}

Considerable evidence points toward a connection between the activity of
quasars and the formation of spheroids.

High resolution imaging has revealed the presence of host galaxies around 
a number of quasars (see e.g. McLure et al. 1999). 
McLeod (1996) finds a correlation between the luminosities of quasars and 
those of the host galaxies.

Both the space density of bright quasars and the total star formation 
rate rise by more than a factor of ten in the interval $0<z<2$ 
(see e.g. Richstone et al. 1998; and Madau 1999), suggesting a common
pattern for quasars and galaxies.
However, the space density of bright quasars drops down earlier
than the total star formation rate and is closer to 
the star formation history of elliptical galaxies, the stellar population
of which is known to be old (see e.g. Renzini 1999).

Hierarchical models of the formation of structures can explain the decline in 
the density of bright quasars at $z\gsim 3$ if it is assumed that quasars
are linked to the 
collapse of the first dark matter [DM] haloes of galactic mass (see e.g. 
Efstathiou \& Rees 1988; and Haehnelt \& Rees 1993).
When
the mass function of DM haloes (given by the Press-Schechter 
formula) is combined with a prescription to associate a galactic luminosity
and a quasar light curve to a DM halo, these models can reproduce
both the luminosity functions of high redshift galaxies and high redshift 
quasars  (see e.g. Haehnelt, Natarajan \& Rees 1998). The main reason 
for restricting this approach to high redshifts is that at low redshift 
there is probably a transition from a phase in which quasars form in gas-rich
nuclei to a phase in which quasars are refuelled by interactions of 
increasingly gas-poor galaxies
(see e.g. Cavaliere, Perri \& Vittorini 1997).

As it is widely believed that the activity of quasars is powered 
by accretion onto a supermassive black hole [BH] 
(see Rees 1984 and references therein),
the detections of supermassive BHs in 
a number of nearby galaxies and the discovery of a correlation between the
mass of the central BH and the luminosity of the host bulge provide
substantial support for the hypothesis that most galaxies contain a 
supermassive BH (see e.g. Ford et al. 1998; Ho 1998; Magorrian et al. 
1998; Richstone et al. 1998; and van der Marel 1998).
A recent study by Salucci et al. (1999a) shows that the mass function of 
supermassive BHs inferred from the relation of BH mass to 
bulge luminosity is consistent with the luminosity function of quasars.
In a subsequent work, Salucci et al. (1999b) find that supermassive BHs
harboured by spiral galaxies are less massive than those detected in 
ellipticals and conclude
that the population of bright quasars is dominated
by objects that form in spheroids.

A possible explanation of the link between quasars and spheroids is that
there is a unifying mechanism fuelling supermassive BHs while
forming spheroids. A mechanism with this
characteristics is known to exist and is provided by mergers.
N-body simulations have shown that mergers of galaxies of comparable masses
(major mergers)
result in the formation of elliptical galaxies (see e.g. Barnes 1988; 
Hernquist 1992, 1993; Hernquist, Spergel \& Heyl 1993; and Heyl, Hernquist \& 
Spergel 1994). Kauffmann \& Charlot (1988) have proven that the merger scenario
for the formation of elliptical galaxies is consistent with
the colour-magnitude relation and its redshift evolution. 
Hydrodynamic simulations
have found that shocks due to mergers can cause a fraction of the gas in the
interacting galaxies to fall at the centre of the merged system
and to fuel a burst of central
activity (Negroponte \& White 1983; Barnes \& Hernquist 1991, 1996; and
Mihos \& Hernquist 1994). These computational results have their observational
counterparts, since a significant fraction of the imaged quasar populations
is known to reside in interacting systems
(see e.g. Bahcall et al. 1997; and McLure et al. 1999).

Cattaneo, Haehnelt \& Rees (1999) have shown that a very simple model in which
after each major merger the central BHs in the progenitors coalesce and
a fraction of the cold gas in the merger remnant is accreted by the merged
BH can reproduce the observed relation of BH mass to bulge luminosity.
Here we investigate whether this model 
can account for the redshift evolution of the quasar population, particularly
at low redshift, where the emission is likely to be dominated by reactivation
of supermassive BHs which have accreted most of their mass at the peak
of the quasar epoch. 
Our results derive from Monte Carlo simulations of the merging histories of
DM haloes in the extended Press-Schechter formalism combined with a
simple scheme for galaxy formation, which has been tested against the total
luminosity function of galaxies, the luminosity function of early type 
galaxies, the redshift
dependence of the total star formation rate and the redshift dependence of
the abundance of neutral hydrogen in damped Lyman-$\alpha$ clouds.
The structure of the paper is as follows.
In Section 2 we describe the computational algorithm used to simulate the
formation of galaxies and the accretion history of supermassive BHs.
In Section 3 we present the results of our calculations. 
Section 4 contains the conclusions of the article.  

\section{Monte Carlo simulations of the formation and evolution of galaxies
and quasars}

\subsection{Merging histories of dark matter haloes}

The standard paradigm for the formation of structure in the Universe
is the gravitational instability of density fluctuations in a primordial
Gaussian random field. In this picture, the number density of collapsed 
DM haloes is accurately described by the Press-Schechter formula 
\begin{eqnarray}
&N(M,z){\rm\,d}M=\\ \nonumber
&-\left({\overline{\rho}\over M}\right)
({1\over 2\pi})^{1\over 2}\left({\omega\over\sigma}\right)
\left({1\over\sigma}{{\rm d}\sigma\over{\rm d}M}\right)
{\rm exp}\left(-{\omega^2\over 2\sigma^2}\right){\rm\,d}M\\ \nonumber
\end{eqnarray}
(Press \& Schechter 1974).
Here $\overline{\rho}(z)$ is the cosmological density at redshift $z$, 
$\sigma_0(M)^2$ is the variance of the linearly extrapolated density field 
on the scale $M$, 
$\omega$ is defined as $\omega(z)\equiv\delta_{\rm c} D(0)/D(z)$,
where $\delta_{\rm c}$ 
is the over-density threshold at which density fluctuations collapse,
and $D(z)$ is the linear growth factor of density fluctuations.  
The variance of the density field is related to the power spectrum of the 
density fluctuations. 
Here we assume a standard cold dark matter power spectrum as given by 
Bond \& Efstathiou (1984), with a normalisation that reproduces  the
present-day space density of clusters of galaxies (Eke, Cole \& Frenk 1996). 
The dependence of the linear growth factor on redshift is given by 
Heath (1977). Here we restrict our attention to the case of a flat
$\Lambda$CDM universe with $\Omega_{\rm M}=0.2$, $\Omega_\Lambda=0.8$ and
a Hubble constant of $75{\rm\,km\,s^{-1}\,Mpc^{-1}}$.

Techniques for generating Monte Carlo realizations of the merging
histories of DM haloes based on extensions of the Press-Schechter 
formula have been described for instance
by Cole \& Kaiser (1988), Cole (1991),
Lacey \& Cole (1993), Kauffmann \& White (1993) and Somerville \&
Kolatt (1999). The basic  
equation is the conditional probability 
that a halo of mass $M$ at redshift $z$ has a progenitor of 
mass $M-{\Delta M}$ at redshift $z+\Delta z$,
\begin{eqnarray}
&P(M\rightarrow M-\Delta M,z\rightarrow z+\Delta z)=\\ \nonumber
&\frac{1}{\sqrt{2\pi}}\;
{\omega(z+\Delta z)-\omega(z)\over[\sigma_0^2(M-\Delta
M)-\sigma_0^2(M)]^
{3\over 2}}\; \exp{\biggl  \{ -{[(\omega(z+\Delta z)-\omega(z)]^2\over
2[\sigma_0^2(M-\Delta M)-\sigma_0^2(M)]} \biggr \}} \\\nonumber
\end{eqnarray}
(Lacey \& Cole 1993).  We follow the  description of 
Somerville \& Kolatt (1999) to construct Monte Carlo realizations
of the merging histories of DM haloes from equation (2). 
The probability distribution (2) is used to assign a mass $M$ to 
a progenitor at redshift $\Delta z$ of a DM halo 
of mass $M_0$ at $z=0$.  The  procedure is 
iterated and further progenitors are drawn from the distribution (2), 
but the merging history  is followed only for haloes with circular 
velocities $v_{\rm c}=({\rm G}M/r_{\rm vir})^{1/2}$
above a certain threshold $v_l$. 
Haloes with $v_{\rm c} <v_{\rm l}$ are treated as accreted mass.
Progenitors with a mass larger than the not yet allocated mass are discarded. 
We construct progenitors of progenitors until all haloes have 
$v_{\rm c} <v_{\rm l}$.   
The procedure is repeated for a representative sample of haloes at $z=0$.   
We adopt a step of $\Delta z=0.01$ and a resolution of 
$v_{\rm l}=70{\rm\,km\,s}^{-1}$.

\subsection{The galaxy formation scheme}

The modelling of galaxies within hierarchal cosmogonies has reached 
a considerable level of complexity. Here we concentrate on the 
effects of mergers on the growth of supermassive BHs. 
Therefore we adopt a
rather simple scheme  for galaxy formation similar to that proposed
by  White and Rees (1978), which should nevertheless catch the
essential features of the hierarchical merging of galaxies (White
1996). 

Important conditions for the 
formation of a galaxy in a collapsed DM halo are
the ability of the gas to cool and the ability of the halo 
to retain its gas in spite of the input of energy and momentum
due to star formation and supernova explosions. 
The ability of the gas to cool depends on the temperature and  the density of 
the gas. For cooling by Bremsstrahlung these  dependencies conspire to give
an upper limit for the mass of the gas that is able to cool efficiently 
(Silk 1977; Rees \& Ostriker  1977). 
Feedback from star formation is more important in haloes with shallower
potential wells and therefore smaller circular velocities
(see i.e. Kauffmann, White \& Guiderdoni 1993). 
Here we do not treat the cooling and feedback explicitly. Instead, we 
introduce an effective efficiency for galaxy formation, which depends  
on the halo circular velocity  in such a way that the mass of the galaxy
associated with a 
DM halo of mass $M_{\rm halo}$ at redshift $z$ scales as 
\begin{equation}
M_{\rm gal}(M_{\rm halo})= \epsilon_{\rm gal}(1+z)^\alpha
M_{\rm halo}  v_{\rm c}^{2}\exp{[-(v_{\rm c}/v_{\rm max})^4]}.
\end{equation}
The parameters $\epsilon_{\rm gal}$ and $v_{\rm max}$ are chosen in such a way
that the luminosity function of our simulated 
galaxies reproduces the observed luminosity function of galaxies
at $z=0$ whereas
$\alpha$ models the redshift dependence of the efficiency of
cooling and is set by the cosmological evolution of the mean 
comoving density of neutral hydrogen in damped Lyman-$\alpha$ absorbers.
The exponential cut-off at high virial velocities mimics the  inability 
of the gas to cool and form stars in the very deep potential wells
which form at late times.

\begin{figure}
\centerline{\psfig{figure=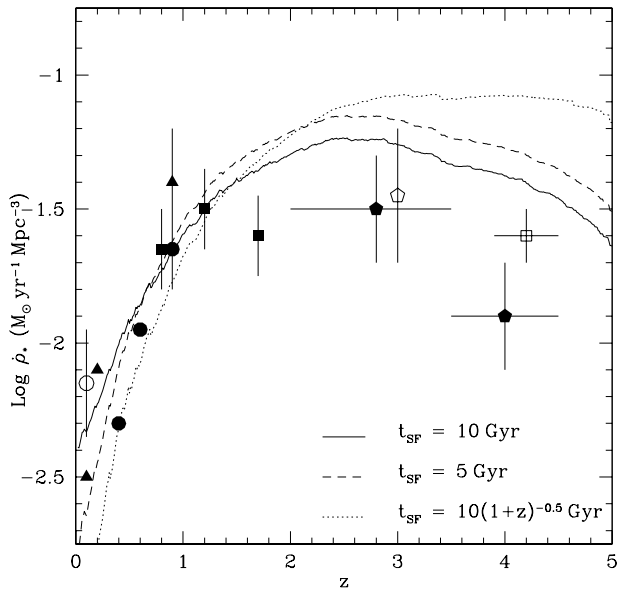,width=0.48\textwidth,angle=0}}
\caption{Mean comoving density of star formation as a function of redshift.
The lines show the results of our simulations for a star formation time-scale
in disks of : i) 10\,Gyr ({\it solid line}); ii) 5\,Gyr ({\it dashed line});
iii) $10(1+z)^{-0.5}\,$Gyr ({\it dotted line}).
The data points have been inferred from the UV-continuum measurements by
Lilly et al. (1996; {\it filled dots}), Connolly et al. (1997;
{\it filled squares}), Madau, Pozzetti \& Dickinson (1998; {\it filled 
pentagons}), Treyer et al. (1998; {\it empty circle}) and Steidel et al.
(1998; {\it empty square}).
The filled triangles show the H$\alpha$ determinations by Gallego et al. 
(1995),
Tresse \& Maddox (1998) and Glazebrook et al. (1998). 
The empty pentagon gives the SCUBA lower limit (Hughes et al. 
1998).}
\end{figure}
\begin{figure}
\centerline{\psfig{figure=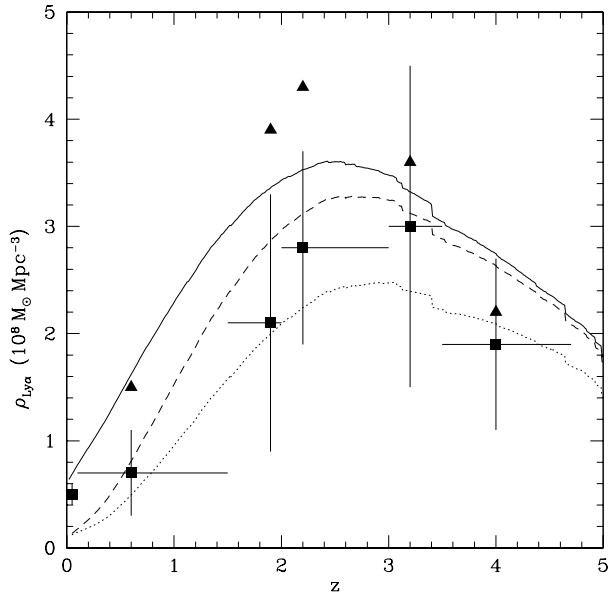,width=0.48\textwidth,angle=0}}
\caption{Mean cosmological density of neutral hydrogen in potential wells
of $v_{\rm c}>70{\rm\,km\,s}^{-1}$ for a star formation time-scale 
in disks of : 
i) 10\,Gyr ({\it solid line}); ii) 5\,Gyr ({\it dashed line});
iii) $10(1+z)^{-0.5}\,$Gyr ({\it dotted line}).
The parameter $\alpha$ in equation (3) sets the peak in the mean
cosmological density of neutral hydrogen.
A peak at $z\sim 2.5$ is obtained for $\alpha=1$.
The data points ({\it filled squares}) give the determinations of the
cosmological density of neutral hydrogen in damped Lymam-$\alpha$ absorbers
by Storrie-Lombardi, McMahon \& Irwin (1996) corrected for possible dust 
obscuration ({\it filled triangles}) and are shown for comparison.
Assuming that most of the cold gas at $z\lsim 4$ is in clouds of 
very high column 
density because dissipation is very effective in concentrating baryons
at the centre of DM haloes, we find 
the best fit to the data on damped Lyman-$\alpha$ absorbers
for $\epsilon_{\rm gal}=0.014$.}
\end{figure}
\begin{figure}
\centerline{\psfig{figure=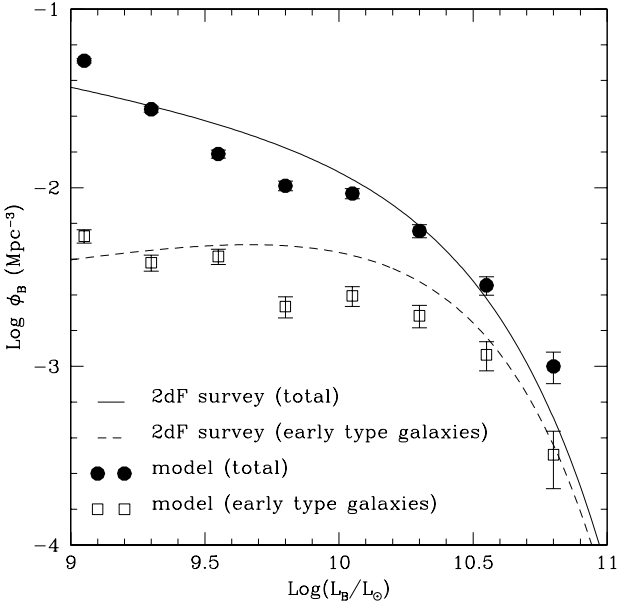,width=0.48\textwidth,angle=0}}
\caption{Observed and simulated  B-band galactic luminosity functions.
Early type galaxies are classified as
Type 1 galaxies in the 2dF redshift survey (Folkes et 
al. 1999). 
A relation of star forming mass 
to  circular velocity and mass of the DM halo of the form 
$M_\star(M,z)= f_\star M_{\rm gal}(M,z)=
0.007(1+z) M_{\rm halo} \times (v_{\rm c}/200\kms) ^{2} 
\exp{[-(v_{\rm c}/300\kms)^4]}$ is assumed. 
Major mergers result in the formation of a bulge 
for mass ratios above $f_{\rm e} = 0.3$.
Stellar masses are converted into B-band luminosities using the K-band
Tully-Fischer relation in de Grijs \& Peletier (1999) 
and the colour corrections in Poggianti (1997).
The total luminosity function and the luminosity
function of early type galaxies are both strongly dependent on the dynamical
friction parameter $\epsilon_{\rm df}$ (equation 4).
An acceptable fit of the data is obtained for $2<\epsilon_{\rm df}<2.5$
and this figure has been plotted for $\epsilon_{\rm df}=2.25$.
Larger values of $\epsilon_{\rm df}$ make mergers less frequent, 
reduce the fraction of elliptical galaxies,
decrease the number of bright galaxies and cause the faint end of the total 
luminosity function to rise.}
\end{figure}

When DM haloes merge, it will take some time before the corresponding
galaxies sink to the centre of the merged halo due to dynamical
friction.  To model this, 
we follow Kauffmann, White \& Guiderdoni (1993)
and assume that initially a single galaxy forms at
the centre of each halo. When haloes merge, the central 
galaxy of the largest progenitor halo becomes the central galaxy
of the new halo. The ``satellite''  galaxies are assumed to  
merge with the central galaxy on the dynamical friction time-scale 
\begin{equation}
t_{\rm df}={1.17\epsilon_{\rm df}
r_{\rm vir}^2v_{\rm c}\over{\rm ln}\left({M/M_{\rm sat}}\right)
{\rm G}M_{\rm sat}}
\end{equation}
(Binney \& Tremaine 1987).
Here $M$, $r_{\rm vir}$, $v_{\rm c}$ are the mass, the virial radius and 
the circular velocity of the new  halo, $M_{\rm sat}$ is the total mass of
the satellite including its DM halo and $\epsilon_{\rm df}$ is a factor that
keeps into account the increase in the time of orbital decay due to tidal
stripping. Recent numerical work by Colpi, Mayer \& Governato (1999) has shown 
that $\epsilon_{\rm df}\gsim 2$ for cosmologically relevant situations.
The mass of the merged galaxy is assumed to 
be the maximum of the sum of the masses 
of the merging galaxies and of the mass given by equation (3). 
If the latter is larger, we interpret the difference  
$\Delta M_{\rm gal}$ as accretion due to inflow from the intergalactic medium
and we add this mass to the total amount of cold gas in the galactic disk.

As in Kauffmann, White \& Guiderdoni (1993), 
we introduce a critical mass ratio $f_{\rm e}$
above which a merger produces an elliptical galaxy. 
For mass ratios below the threshold the smaller galaxy is 
assumed to be tidally disrupted and its mass is added
to the disk of the larger one. 
The value of $f_{\rm e}$ is determined 
by requiring the Monte Carlo simulations to reproduce the 
observed bulge luminosity function, 
once the other free parameters have been fixed.

We model star formation by assuming that 
only a fraction $f_\star$ of the mass in cold gas forms 
visible stars and that
the conversion of cold gas into stars
is described by a purely exponential law with a time-scale of 1\,Gyr for
elliptical galaxies and 10\,Gyr for spiral galaxies.
We have also tried other models, 
but the results have been less satisfactory (Figures 1
and 2).
The initial reservoir of cold gas available to form stars
is comprised by a fraction $f_\star$ of the total amount of gas 
cooled onto the galaxy since the last major merger for a spiral galaxy 
and by a fraction $f_\star$ of
the total amount of cold gas present in the disks of the parent galaxies
for an elliptical galaxy. 

To convert the galactic mass function generated from Monte Carlo simulations
into a $B$-band luminosity function, we solve
equation (3) in $v_{\rm c}$ and substitute the solution into 
the $K$-band Tully-Fisher relation to derive the $K$-band luminosity for a
galaxy of given $M_{\rm gal}$ and $f_\star$.
The $K$-band luminosity is then converted into a $B$-band luminosity using the
colour correction appropriate for the galactic morphological type. 
In our calculations we use
the $K$-band Tully-Fisher relation in
de Grijs \& Peletier (1999) and the $B-K$ corrections in Poggianti (1997). 

Figures 1 and 2 show the redshift dependence of 
the total mean density of
star formation and of the mean cosmological mass density in cold gas as they 
result from our Monte Carlo simulations.
The total mean star formation density is dominated by star formation in
disks even if about half of the galactic mass ends up in spheroids
because mergers convert disk stars into bulge stars. 
We fit the dust-corrected mean cosmological density of neutral hydrogen
in damped Lyman-$\alpha$ absorbers for a cooling efficiency  
of $\epsilon_{\rm gal}\gsim 0.014$. The equality holds if 
dissipation is very effective in concentrating baryons at the centre of
DM haloes and most of the mass in neutral hydrogen at $z\lsim 4$ is
in clouds of very high column density.
For the mass-to-light ratio inferred from the Tully-Fisher relation,
this high value of $\epsilon_{\rm gal}$ can be reconciled with the
bright end of the luminosity function of galaxies
if the fraction of cold gas that goes into star formation is 
not much more than a half.
The remainder must be either reheated and ejected or
end up in a population of sub-stellar
objects, which can account for the dark matter in disks and baryonic haloes.
Since it is quite improbable that a galaxy with a stellar mass of
$\sim 10^{11}M_\odot$ has lost half of its baryonic mass through ejection
by supernovae, it is a likely conclusion of 
our model that at least in massive DM
haloes a significant fraction of the mass in cold gas has to go into
a population of sub-stellar objects.
Figure 3 compares the luminosity functions resulting from our simulations to
those observed. 
If we assume a cooling efficiency of $\epsilon_{\rm gal}=0.014$, we
find the best fit to the observations for $f_\star=0.53$.

\subsection{The growth of black holes through mergers and accretion}

Supermassive BHs grow through merging with other supermassive BHs and through 
accretion of gas. They may also grow through tidal disruption
and accretion of stars, but we do not want to discuss this
possibility here.
 
After a galactic merger has taken place, the supermassive BHs contained
in the merging galaxies sink at the centre of the new galaxy, form a binary
system and eventually coalesce.
The time-scale for this process in uncertain and its final outcome can also
be quite different, if another galactic merger intervenes before the 
supermassive BHs have time to merge (Begelman, Blandford \& Rees 1980),
but here we assume that this time-scale is much shorter than the Hubble time,
so that the supermassive BHs contained in the merging galaxies always
merge. 

The amount of cold gas accreted by the supermassive BH will have a
complicated dependence on halo and galactic properties (dynamical and
kinematical structure, abundance of cold gas, past merging history,
angular momentum),
and will ultimately depend on the physics of the mechanism responsible
for driving the gas to the centre,
but here we concentrate on a very simple model where after each major merger a
fraction of the gas in the
merger remnant falls to the centre and fuels the BH on a time-scale of 
$10^7\,$yr. The mass accretion rate is the minimum between the rate 
of fuel supply and the Eddington limit:
\begin{equation}
\dot{\cal M}_\bullet={\rm min}\{{M_{\rm res}\over 10^7{\rm\,yr}},
{M_\bullet\over\epsilon_{\rm rad}t_{\rm S}}\},
\end{equation}
where $\dot{\cal M}_\bullet$ is the rest mass accretion rate,
$M_{\rm res}$ is the mass in the gas reservoir of the supermassive BH,
$M_\bullet$ is the mass of the central BH,
$\epsilon_{\rm rad}$ is the efficiency at which the accreted rest mass
is converted into radiation and $t_{\rm S}$ is the Salpeter time.
The evolution of $M_\bullet$ is computed from $\dot{\cal M}_\bullet$ 
assuming that gas is accreted from the innermost stable circular orbit
(see e.g. Bardeen 1970).

Equation (5) implies that a BH cannot power a bright quasar if it has not 
reached a certain mass, 
even if the fuel supply is potentially available. 
Meanwhile, star formation is competing with the BH in depleting its gas 
reservoir.
The time over which a BH reaches the mass required to power a bright quasar
depends on the initial mass of the BH and on the efficiency for converting
accreted mass into radiation. Figure 4 shows the initial BH mass required
for a bright quasar to form on a time-scale shorter than the star formation
time-scale as a function of the radiative efficiency.
From this figure we see that, for interesting values of the radiative 
efficiency and a star formation time-scale $>10^8\,$yr,
the central BH forms and evolves into a bright quasar independently of the 
initial mass and therefore of whether the BHs in the merging galaxies
have coalesced or not. If, instead,
the star formation time-scale in the central region of the merger remnant
is $\lsim 10^8\,$yr, then it becomes relevant to know what the seed mass is and
how frequently supermassive BHs merge.
In this article we restrict ourselves to
the first possibility to avoid the introduction of
further free parameters which might be difficult to constrain.

\begin{figure}  
\centerline{\psfig{figure=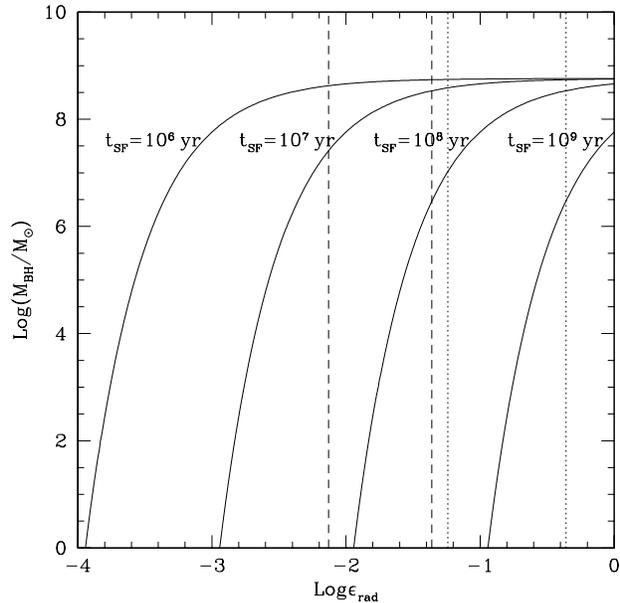,width=0.48\textwidth,angle=0}}
\caption{Minimum initial mass
for a BH to evolve into a bright quasar ($M_{\rm B}<-26$) 
over a time shorter than the star formation time-scale
as a function of the radiative efficiency $\epsilon_{\rm rad}$.
A quasar radiating at the Eddington limit has a blue magnitude
$M_B=-26$ for a mass of $\sim 6\times 10^8\,M_\odot$ and a $B-$band
bolometric correction of 0.087.
Curves have been plotted for a star formation time-scale of 
$10^6$, $10^7$, $10^8$ and $10^9$ years ({\it solid lines}).
The dashed lines enclose the observationally plausible interval of values
for $\epsilon_{\rm rad}$ determined by comparing the mass density in
supermassive BHs inferred from counts of unobscured optically selected quasars
(Choski \& Turner 1992)
to the density derived from the estimated masses
of supermassive BHs in the nuclei of nearby galaxies
(Magorrian et al. 1998; Franceschini, Vercellone \& Fabian 1999;
and Salucci et al. 1999a).
The dotted lines show the maximum values of $\epsilon_{\rm rad}$ permitted in
the Schwarzschild ($\epsilon_{\rm rad}=0.057$) and in the Kerr 
($\epsilon_{\rm rad}=0.432$) metric.}                                  
\end{figure}

\section{The redshift evolution of the quasar population}

Carlberg (1990) speculated that the 
the decline in the density of bright quasars after
the peak of the quasar epoch
might be explained purely in terms of a decrease in the frequency of mergers.
For a ratio of BH mass to bulge mass of 0.006 (Magorrian et al. 1998)
and a $B-$band bolometric correction for quasars of 0.087 (Pei 1995), 
bright quasars of $M_{\rm B}<-26$ and $M_\bullet\gsim 6\times 10^8\,M_\odot$ 
are associated with large ellipticals of $M_{\rm gal}\gsim 10^{11}M_\odot$
and the quasar epoch should be marked by the epoch at which large ellipticals 
are assembled. However,
the frequency of major mergers forming ellipticals of 
$M_{\rm gal}>10^{11}M_\odot$ (Figure 5, solid line) does not trace the redshift
evolution of the comoving density of bright quasars (Figure 6) 
because the decrease in the frequency of mergers is compensated by 
an increase in the number of bright galaxies at low redshift.
The conclusion is that the decline in the merging rate is not sufficient to
explain the fall in the comoving density of bright quasars at $z\lsim 2$.

The key element which is
missing in this analysis is the consumption of cold gas by
star formation resulting in increasingly gas-poor mergers.
In our simulations we reproduce a ratio of BH mass to bulge mass 
of 0.006 if we assume that after each major merger the central BH accretes
$\sim 2\%$ of the cold gas in the merging remnant.
\begin{figure}  
\centerline{\psfig{figure=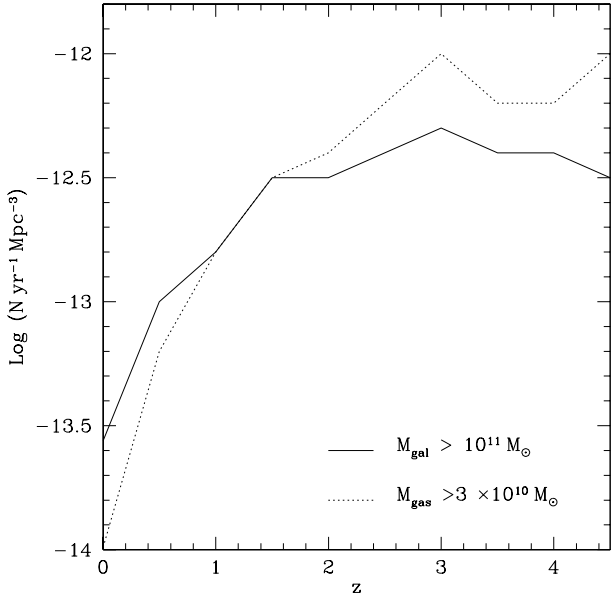,width=0.48\textwidth,angle=0}}
\caption{The redshift evolution of the merging rate for mergers of galaxies 
of comparable masses (mass ratio $f_{\rm e}>0.3$) containing
a total baryonic mass larger than $10^{11}M_\odot$ ({\it solid line}) and 
a mass in cold gas larger than $3\times 10^{10}M_\odot$ ({\it dotted line}).
The merging rates are computed for a dynamical friction parameter of 
$\epsilon_{\rm df}=2.25$ (see Section 2.2 and the caption to Figure 3).}
\end{figure}
\begin{figure}  
\centerline{\psfig{figure=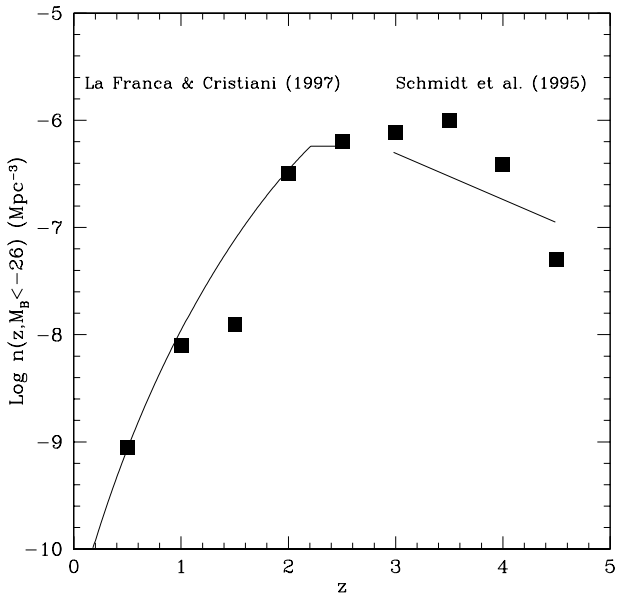,width=0.48\textwidth,angle=0}}
\caption{Redshift evolution of the comoving number density of bright quasars
($M_{\rm B}<-26$). The filled squares are the results of the 
simulations
for an accreted mass after each major merger of 
$M_{\rm accr}={\rm min}(3\times 10^{-4}(1+z)^3, 0.0125)M_{\rm gas}$,
an accretion time-scale of $2\times 10^7\,$yr
and an efficiency for converting accreted mass into blue light of 
$\sim 3.8\times 10^{-3}$.
The solid lines show the evolution
inferred from the luminosity 
function in La Franca \& Cristiani (1996) and a fit to the counts of bright
high redshift quasars in Schmidt et al. (1995).}                              
\end{figure}
\begin{figure}  
\centerline{\psfig{figure=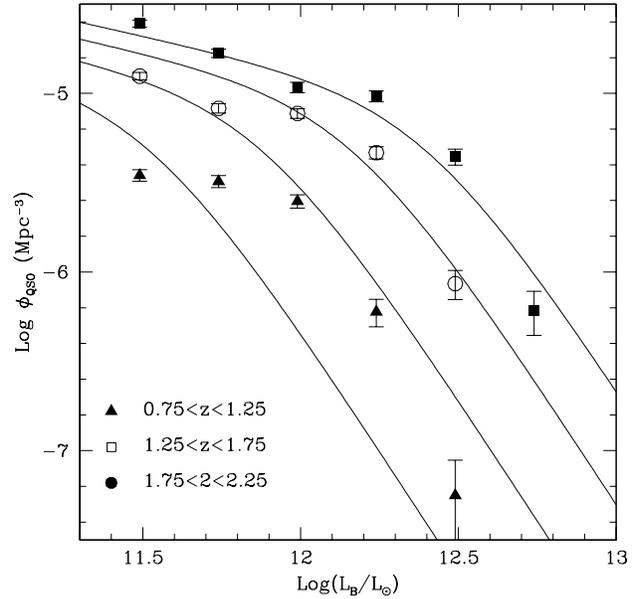,width=0.48\textwidth,angle=0}}
\caption{Evolution of the quasar luminosity function between $z\sim 2$
and $z\sim 1$.
The points show the results of the simulations
for the model described in the caption of Figure 6. 
The solid lines give fits to the data for $z=0.75,\,1.25,\,1.75,\,2.25$  
derived by La Franca \& Cristiani (1996) assuming a 
a double power law and a pure luminosity evolution.} 
\end{figure}
\begin{figure}  
\centerline{\psfig{figure=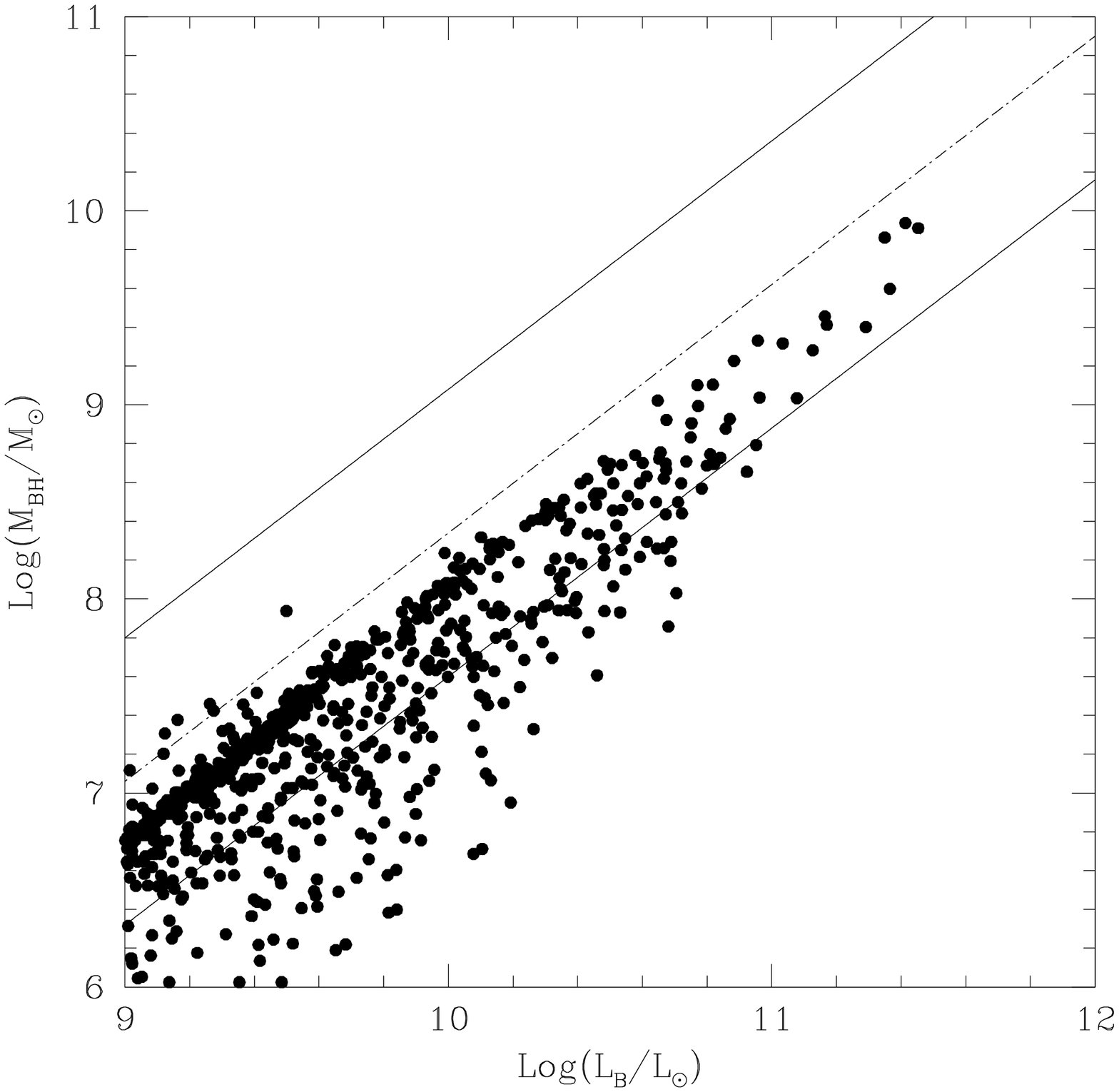,width=0.48\textwidth,angle=0}}
\caption{BH masses and bulge luminosities for an accreted mass after
each major merger of
$3\times 10^{-4}(1+z)^3$ the mass of the cold gas in the merger remnant. 
This fraction increases with redshift up to
a maximum of 1.25\%. The introduction of this maximum value prunes points
above the linear regression (some scatter survives due to spheroids
which have suffered tidal disruption). The model shown in this figure gives
a present mean mass density in supermassive BHs of 
$\sim 10^6M_\odot{\rm Mpc}^{-3}$.
The solid lines enclose the region within $\pm 1\sigma$ in
the linear least square fit to the data on the masses of 
supermassive BHs in nearby
galaxies as compiled by Ho (1998) and Magorrian et al. (1998).
See also Cattaneo, Haehnelt \& Rees (1999).}   
\end{figure}
We can therefore refine our analysis considering the merging rate for
galaxies with
a total mass in cold gas larger than  $3\times 10^{10}M_\odot$ (Figure 5,
dotted line). The frequency of gas-rich mergers shows a stronger evolution
than the frequency of mergers involving a large total mass, but the extent of
this evolution is still insufficient to explain the decline of two orders of
magnitude in the density of bright quasars between the peak of the quasar 
epoch and $z\sim 1$. 

To explain the redshift evolution of the density of bright quasars in the
interval $0<z<2$ we need to assume a dependence on redshift either in the
fraction of available gas accreted or in the time-scale for accretion.
The results of our simulations are consistent with counts of bright quasars
and
observations of the quasar luminosity function for
an accreted mass after each major merger of 
$M_{\rm accr}={\rm min}(3\times 10^{-4}(1+z)^3, 0.0125)M_{\rm gas}$,
an accretion time-scale of $2\times 10^7\,$yr
and an efficiency for converting accreted mass into
blue light of
$\epsilon_{\rm rad}\sim 3.8\times 10^{-3}$
(Figures 6 and 7).
Here $M_{\rm gas}$ is the mass of cold gas in the merging remnant. 
The scaling as $(1+z)^3$ has been introduced to mimic
a proportionality of the fraction of accreted
gas to the density of the host galaxy.
The upper limit to the fraction of cold gas that can be accreted 
has been set to prevent unphysical situations where
most of the gas in the merging remnant ends up in the central BH.
The introduction of this upper limit affects the relation of BH mass to
bulge luminosity, since it prunes most of the point above a certain ratio
of BH mass to bulge luminosity (Figure 8).
There is no evidence for such a sharp
upper limit in the data (see e.g. Ho 1998; Magorrian et al. 1998; and
Cattaneo, Haehnelt \& Rees 1999), but this could simply mean that there is
an intrinsic scatter in the maximum fraction of gas that can be accreted by
the supermassive BH. The origin of this scatter could be found e.g. in a
dependence on the angular momentum of the DM halo, 
in the dynamics of the merger
(impact parameter, eccentricity, morphologies of
the merging galaxies, etc.) or in the complicated hydrodynamics responsible
for driving gas into the centre.
If the upper limit is removed, the comoving density of bright quasars
continues rising up to $z\sim 3$ and does not start declining until $z\gsim 4$.
Kauffmann \& Haehnelt (1999) have recently incorporated into semi-analytic
models of galaxy formation a scheme for the growth of supermassive BHs
similar to the one
presented in this article, but containing
a redshift dependence in the accretion
time-scale ($t_{\rm accr}\propto t_{\rm dyn}\propto (1+z)^{-1.5}$) rather
than in the fraction of accreted mass. They find that the steep decline in the
comoving density of bright quasars begins at $z\lsim 1$ rather than at
$z\lsim 2$, 
most likely because a scaling as $(1+z)^{1.5}$ is insufficient
to explain the striking evolution of the quasar population.

\section{Conclusion}

We have investigated a very simple unified model in which both spheroids and 
supermassive BHs powering quasars form through mergers of galaxies of
comparable masses. We have assumed that cooling only forms disk galaxies and
that, whenever two galaxies of comparable masses merge, the merging remnant
is an elliptical galaxy, a burst of star formation takes place and a fraction
of the gas in the merging remnant is accreted by a central supermassive BH
formed by the coalescence of the central BHs in the merging galaxies.

We have found that this simple model
is consistent with the shape of the quasar
luminosity function, but we have also found that its redshift evolution
cannot be explained purely in terms of a decrease in the merging rate and of
a decline in the amount of fuel available. 

To explain the striking evolution
of the space density of bright quasars in the interval $0<z<2$, we need
to make additional assumptions, such as a dependence on redshift either in
the fraction of available gas accreted or in the time-scale for accretion.
With these additional assumptions it is possible to obtain results consistent
with the redshift evolution of the quasar luminosity function and with the
amount of scatter in the relation of BH mass to bulge luminosity
(on this point see also Cattaneo, Haehnelt \& Rees 1999).

\section{Acknowledgements} 

Andrea Cattaneo thanks Martin J. Rees and Piero Madau for useful conversation
and acknowledges the support of an Isaac Newton Studentship.

\end{document}